\documentclass{aastex631}

\usepackage[toc,page]{appendix}

\received{..., 2022}
\revised{..., 2022}
\accepted{\today}
\submitjournal{ApJ}

\shorttitle{}
\shortauthors{Penza et al.}
\graphicspath{{./}{figures/}}

\begin{document}
\title{Investigating the Temperature Sensitivity of UV Line Ratios in the 280 nm Region of Solar-like Stars}

\correspondingauthor{Serena Criscuoli}
\email{scriscuo@nso.edu}

\author[0000-0002-3948-2268]{Valentina Penza}
\affiliation{Dipartimento di Fisica, Universit\`a degli Studi di Roma Tor Vergata, Via della Ricerca Scientifica 1, Roma, 00133, Italy}

\author[0000-0002-4525-9038]{Serena Criscuoli}
\affiliation{National Solar Observatory, 3665 Discovery Dr., Boulder, CO 80303, USA}

\author[0000-0001-8623-5318]{Raffaele Reda}
\affiliation{Dipartimento di Fisica, Universit\`a degli Studi di Roma Tor Vergata, Via della Ricerca Scientifica 1, Roma, 00133, Italy}

\author[0000-0002-1155-7141]{Luca Bertello}
\affil{National Solar Observatory, 3665 Discovery Dr., Boulder, CO 80303, USA}

\author[0000-0002-4896-8841]{Giuseppe Bono}
\affiliation{Dipartimento di Fisica, Universit\`a degli Studi di Roma Tor Vergata, Via della Ricerca Scientifica 1, Roma, 00133, Italy}
\affiliation{INAF -- Osservatorio Astronomico di Roma, via Frascati 33, Monte Porzio Catone, italy}

\author[0000-0003-2500-5054]{Dario Del Moro}
\affiliation{Dipartimento di Fisica, Universit\`a degli Studi di Roma Tor Vergata, Via della Ricerca Scientifica 1, Roma, 00133, Italy}

\author[0000-0002-2662-3762]{Valentina D'Orazi}
\affiliation{Dipartimento di Fisica, Universit\`a degli Studi di Roma Tor Vergata, Via della Ricerca Scientifica 1, Roma, 00133, Italy}
\affiliation{INAF -- Osservatorio Astronomico di Padova, vicolo dell'Osservatorio 5, 35122, Padova, Italy}

\author[0000-0001-7369-8516]{Luca Giovannelli}
\affiliation{Dipartimento di Fisica, Universit\`a degli Studi di Roma Tor Vergata, Via della Ricerca Scientifica 1, Roma, 00133, Italy}

\author[0000-0001-8044-5701]{Giuseppina Nigro}
\affiliation{Dipartimento di Fisica, Universit\`a degli Studi di Roma Tor Vergata, Via della Ricerca Scientifica 1, Roma, 00133, Italy}

\author[0000-0002-2276-3733]{Francesco Berrilli}
\affiliation{Dipartimento di Fisica, Universit\`a degli Studi di Roma Tor Vergata, Via della Ricerca Scientifica 1, Roma, 00133, Italy}




\begin{abstract}
Stellar UV spectra are fundamental diagnostics of physical and magnetic properties of stars.
For instance, lines like \ion{Mg}{2} at 280 nm serve as valuable indicators of stellar activity, providing insights into the activity levels of Sun-like stars and their potential influence on the atmospheres of orbiting planets. On the other hand, the effective temperature ($T_{eff}$) is a fundamental stellar parameter, critical for determining stellar properties such as mass, age, composition and evolutionary status. In this study, we investigate the temperature sensitivity of three lines in the mid-ultraviolet range (i.e., \ion{Mg}{2} 280.00 nm, \ion{Mg}{1} 285.20 nm, and \ion{Si}{1} 286.15 nm). Using spectra from the International Ultraviolet Explorer (IUE), we analyze the behavior of the ratios of their corresponding indices (core/continuum) for a sample of calibrating solar-like stars, and find that the ration R = \ion{Mg}{2}/\ion{Mg}{1} best traces $T_{eff}$ through a log-log relation. The $T_{eff}$ estimated using this relation on a test-sample of solar-like stars agree with the $T_{eff}$ from the literature at  the 95\% confidence level. The observed results are interpreted making use of Response Functions as diagnostics. 
This study extends the well-established use of line depth ratio-temperature relationships, traditionally applied in the visible and near-infrared ranges, to the mid-UV spectrum. With the growing interest in stellar UV spectroscopy, results presented in this paper are potentially relevant for future missions as HWO, MANTIS and UVEX.

\end{abstract}

\section{Introduction} 
\label{sec:intro}


The UV stellar spectral region and individual lines hold great significance in both  stellar and planetary physics. UV studies have revolutionized our understanding of massive stars and their stellar winds \citep[e.g.,][]{Hillier2020}, while UV lines serve as valuable tools for refining physical models of stellar upper atmospheres \citep[e.g.,][]{Loyd2021}. Furthermore, the UV region of stellar spectra is crucial for understanding star-exoplanet interactions. Actually, variations in stellar UV irradiance, linked to magnetic activity, significantly impact exoplanet atmospheres, driving short-term changes in the middle atmosphere and long-term climate shifts, as well as altering their chemical composition via photodissociation and inducing atmospheric erosion processes \citep[see, e.g.,][]{Sanz-Forcada2010, Tilley2019, Reda2023}.
For oxygen-rich planetary atmospheres, as in the case of Earth, the 200-300 nm spectral region is particularly important, as it governs the formation and destruction of ozone, a key component of the stratosphere \citep[e.g.,][]{Bordi2015, lovric2017}. 
Therefore, characterizing the properties of UV lines in relation to stellar parameters, especially for Sun-like stars considered potential hosts for habitable planets, is of fundamental importance. 
The effective temperature ($T_{\rm eff}$) is one of the most important stellar parameters and serves as a pivotal parameter in the study of stellar atmospheres, playing a crucial role in inferring properties such as mass, age, surface gravity, and evolutionary status, along with providing insights into the star's chemical composition. Various techniques exist to determine $T_{\rm eff}$, with the primary method relying on direct calculation of the radius of a star and absolute luminosity. However, this approach is limited to nearby stars. Alternative methods based on color-temperature relations \citep[e.g.,][]{alonso1996,casagrande2021} or high-resolution spectroscopy \citep[see e.g.,][]{gehren1981,cayrel1963,Lind2012} become essential for the more distant ones. Spectroscopic techniques, especially those that focus on stellar absorption lines, provide a more robust avenue for the determination of $T_{\rm eff}$, as they are largely unaffected by interstellar extinction. One notable method is the use of line depth ratios (LDRs), which entails measuring the relative strengths of absorption lines with different excitation potentials. This technique, first introduced by \cite{gray91}, uses the temperature-dependent behavior of these lines to achieve precise temperature estimates. Recent studies have expanded this method across a range of stellar classifications and spectral features, demonstrating its versatility and effectiveness in diverse contexts, including the analysis of variable stars (see \citealt{caccin02,kov2006,kov2023,biazzo2007,Fukue, afsar2023, taniguchi2018, taniguchi, Jian2019, matsunaga}, and references therein). 
The estimate of the effective temperature in radial variables (Cepheids, RR Lyrae) is, indeed, a challenging problem, since typical radial variables experience variations of the order of one thousand degrees along the pulsation cycle, while the surface gravity changes by 0.2/0.3 dex and the micro-turbulent velocity can change from two to four km/s \citep{bono24}. This means that accurate estimates of the atmospheric parameters from high-resolution spectra and, in particular, of the effective temperature are required to provide accurate elemental abundances.




In this paper, we evaluate the temperature sensitivity of three line ratios in the middle ultraviolet range (MUV, 200–300 nm). For this purpose, we used spectra from the International Ultraviolet Explorer (IUE) space telescope, pioneering mission that provided a homogeneous and extensive database of stellar UV spectra. IUE operated continuously for 18 years, from 1978 to 1996, with two spectrographs: the long-wavelength spectrograph in the wavelength range of 185.0 to 330.0 nm (MUV, i.e., middle-UV) and the short-wavelength spectrograph in the range of 115.0 to 200.0 nm (FUV, i.e., far-UV). With the growing interest in stellar UV spectroscopy, several new missions are operational or planned to be launched in the near future, such as CUTE \citep{CUTE}, HWO \citep{HWO}, MANTIS \citep{indahl2024}, MAUVE \citep{MAUVE} and UVEX \citep{kulkarni2021}, including near-UV (NUV) observations.

Although most of the radiative energy in the Sun is distributed mainly in visible areas, it is well known that the UV component contributes the most to the variability of 
the bolometric flux induced by magnetic activity \citep{Yeo2015, berrilli2020, ErmolliI2003, ermolli2013}. The MUV spectral region (180-300 nm), in particular, contains a wide variety of spectral features that may serve as diagnostics for key stellar parameters, such as effective temperature, surface gravity, and metal abundance \citep{Fanelli}. In addition, this wavelength range has been extensively studied due to the presence of lines considered excellent proxies for magnetic chromospheric activity in the Sun and in F-G-K stars, such as \ion{Mg}{2} at 280.0 nm and \ion{Mg}{1} at 285.2 nm \citep{Schrijver92, deland, heath86, viereck2004, buccino08, linsky2017, Criscuoli2018,  kim22}. In particular, the \ion{Mg}{2} h \& k resonance lines (at 280.3 nm and 279.6 nm) are formed in a way similar to the \ion{Ca}{2} H \& K lines (396.8 nm and 393.4 nm), which have been the longest monitored lines for the study of stellar activity, beginning in the 1960s with the HK Project at Mount Wilson Observatory \citep{Wilson1968, Wilson1978}.
For the Sun, disk-integrated emission in the K line of \ion{Ca}{2}, referred to as the \ion{Ca}{2} K index \citep[see, e.g.,][]{Bertello2016}, has been shown to strongly correlate with the \ion{Mg}{2} index, both overall and in their components at time scales longer than the rotational period \citep{RedaPenza2024}.

These UV lines are also fundamental diagnostics of stellar chromospheres \citep[e.g.][]{Houdebine, fontenla2011, linsky2017,tilipman2021,peralta2022,peralta2023}. For example, inversions of high-spatial, high-spectral observations at the \ion{Mg}{2} spectral range obtained with IRIS \citep{depontieu2014}, allows us to estimate the chromospheric stratification of different features observed in the solar chromosphere \citep[e.g.][]{sainzdalda2019,dasilvasantos2020, jev2022}. 

Here, we take into account the \ion{Mg}{2} and \ion{Mg}{1} lines and also the \ion{Si}{1} 288.16 nm line \citep{morrill}. 
We compute the ratios of the corresponding indices (core/continuum) for a sample of calibrating solar-like stars and investigate their relationship with $T_{eff}$. The observed trends are analyzed and justified by evaluating the corresponding Response Functions (RFs) of the ratios. RFs enable us to quantify the sensitivity of emergent intensity to small perturbations in the thermodynamic parameters across different depths of the stellar atmosphere \citep{caccin77, landi77, ruizcobo94, milic2017}.
Finally, using the established relationship between \ion{Mg}{2}/\ion{Mg}{1} and $T_{eff}$, we determine the effective temperature values for a second sample of solar-like stars.

\section{The data and line index ratios} \label{sec1}

\subsection{The stellar dataset}
We consider the UV spectra dataset available from the IUE public library\footnote{\url{https://archive.stsci.edu/iue/search.php}}. We have selected stars that posses the following characteristics:

\begin{itemize} 

\item Effective temperature and gravity in the ranges $\mathrm{5000 < T_{eff} (K) < 6500}$ and $\mathrm{4 < log \, g (dex) < 5}$ (considering the individual random errors);

\item The presence in the database of more than three high-dispersion spectra, which allows us to calculate average values and reduce the effects of any intrinsic variations due to magnetic activity;

\item The presence in the spectra of the three lines \ion{Mg}{2}, \ion{Mg}{1}, and \ion{Si}{1}, which are found in the spectral range between 276.5 and 288.5 nm. 

\end{itemize}

\begin{figure}[h]
\centering
\includegraphics[scale=0.65]{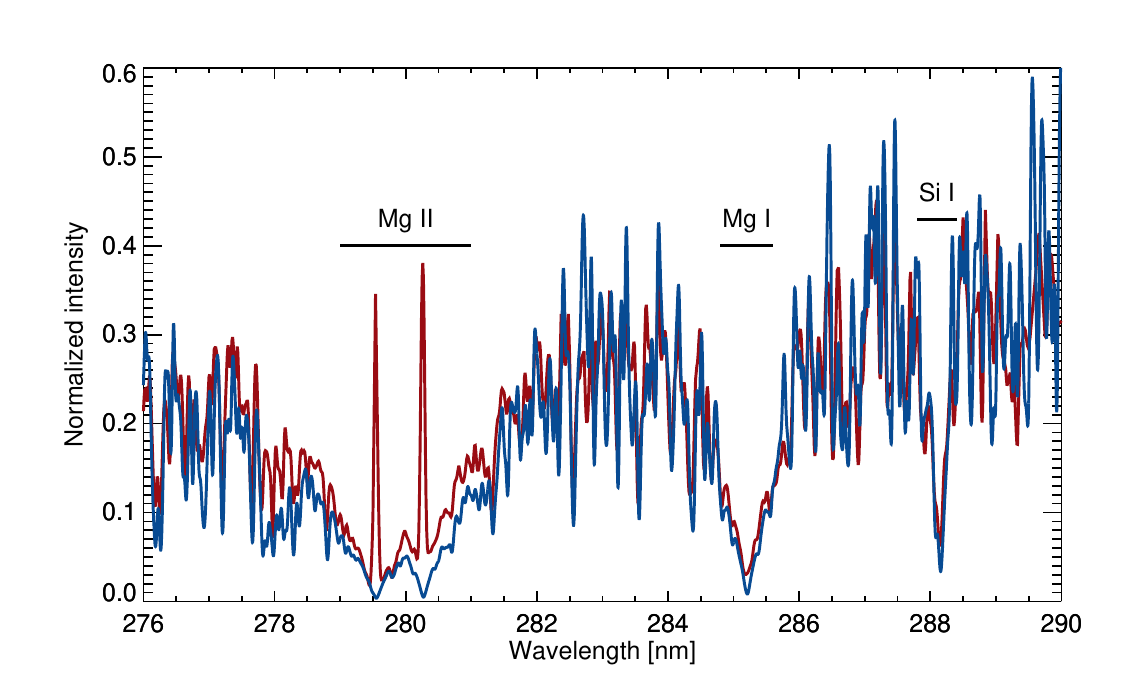}
\caption{ 
Observed spectrum of HD 20630 (red line) in the spectral region utilized in our investigation. A synthetic spectrum (blue line), generated using the SPECTRUM code with parameters $T_{eff}$ = 5750 K, log g = 4.5, and solar metallicity [Fe/H], is also displayed. This synthetic spectrum was smoothed using a Gaussian kernel with a width of 0.02 nm. The black horizontal segments represent the integration range for the cores of the respective lines reported in Tab.~\ref{tab2}}. 
\label{spectrum}
\end{figure}

The 52 stars selected using the above criteria are listed in Table \ref{tab1}, where we provide the values of their effective temperature ($\mathrm{T_{eff}}$), B-V color, log g, $\mathrm{[Fe/H]}$\footnote{We adopt the standard spectroscopic notation such that $\text{[Fe/H]} = \log_{10} \left( \frac{N_{Fe}}{N_{H}} \right) - \log_{10} \left( \frac{N_{Fe, \odot}}{N_{H, \odot}} \right)$}, $\mathrm{[\alpha/Fe]}$, age, rotation period ($P_{rot}$) and the S-index, where available. In Fig.~\ref{spectrum}, we provide an example of the HD 20630 spectrum, along with a synthetic spectrum degraded to IUE spectral resolution (0.02 nm), for comparison. The synthetic spectrum was computed using the SPECTRUM program \footnote{\url{https://www.appstate.edu/~grayro/spectrum/spectrum.html}} by \cite{grayRO94}, under the hypothesis of local thermodynamic equilibrium (LTE), and with an atmospheric model from the Kurucz grid \citep{kurucz}\footnote{\url{http://kurucz.harvard.edu/grids.html}} with $\mathrm{T_{eff} = 5750}$ K, $\mathrm{log\,g = 4.5}$ dex and solar metallicity \citep{grevesse}. 
The values of the atmospheric parameters $\mathrm{T_{eff}}$, $\mathrm{log\,g}$, and $\mathrm{[Fe/H]}$ reported in Table~\ref{tab1} are calculated by averaging all available values for each star within the PASTEL catalogue\footnote{\url{https://vizier.cds.unistra.fr/viz-bin/VizieR?-source=B/pastel}} \citep{Soubiran2016}, with the associated confidence interval obtained as one standard deviation of these values. The B-V values are taken from the SIMBAD\footnote{\url{http://simbad.cds.unistra.fr/simbad/sim-fid}} database, except where explicitly indicated. For the Sun, the values of $\mathrm{T_{eff}}$ and $\mathrm{log\,g}$ are from \cite{melendez}. The reference solar metallicity is based on \cite{Asplund2009}, where the logarithmic number density of iron is reported as $7.50\pm0.04$.

The S-index values are obtained from the Mount Wilson Observatory HK Project catalog \footnote{\url{https://dataverse.harvard.edu/dataverse/mwo_hk_project}}\textbf{\citep{MW}}
except where explicitly indicated. For each star, the S-index measurements were filtered by removing outliers beyond 4$\sigma$. This procedure allows us to exclude extremely large and small values in the datasets that could correspond to spurious measurements or data affected by specific observational problems, which are sometimes present in the Mount Wilson data \citep[see e.g.,][for the case of HD 115383]{DiMauro2022}. 
For the solar case, we utilized the Ca II K index dataset from \citet{Bertello2016}. We calculated the average value and standard deviation over the last five solar cycles (1964–2019) and converted these to the S-index scale using the relationship described by \citet{Egeland2017}.

\LTcapwidth=\linewidth
\begin{longtable}{c|c|c|c|c|c|c|c|c}
\caption{List of the selected stars and their stellar parameters. The B-V values with * are not derived by SIMBAD database. Specifically: HD 35296, HD 115383, HD 154417, HD 187691, HD 206860 by \cite{pizzolato}, HD 82443 by \cite{boro_saikia}, HD 143761 by \cite{choi}, and HD 187013 by \cite{brandenburg}. $\alpha$-element abundances come from the analysis of Gaia RVS spectra by \citet{recioblanco2023}; in this case, the typical uncertainties are around 0.1 - 0.15 dex. The stellar ages are from \citet{Takeda2007}, expect for HD 81809 by \citet{Fuhrmann2018}, HD 115383 by \citet{DiMauro2022}, HD 18256 and HD 115404 by \citet{Mittag2023}. $P_{rot}$ values are from \citet{Baliunas1996}, \citet{Hempelmann2016}, \citet{Brandenburg2017} or \citet{Olspert2018}, except for HD 166 by \citet{Gaidos2000}, HD 1581 by \citet{mascareno}, HD 2151 by \citet{Metcalfe2024}, HD 19994 by \citet{Mayor2004}, HD 20794 by \citet{Pepe2011}, HD 147513 by \citet{Hussain} and HD 187013 by \citet{Saar1999}. The S-index values with * are not derived from the Mount Wilson database. Specifically: HD 2151 and HD 128620A by \cite{buccino08}, HD 20794 by \cite{Basturk2011}, HD 33262 and HD 44594 by \cite{Schroder}, HD 147513 by \cite{Hussain} and HD 1581 by \cite{mascareno}.}

\endfirsthead
\multicolumn{9}{l}{\textit{(continued.)}}
\endhead
STAR & $T_{eff}$ &  B-V  & log g  & [$Fe/H$] &  [$\alpha/Fe$] & Age & $P_{rot}$ &   S-index\\
HD    &   (K)         &       &  (dex) &        &      & (Gyr) & (days) & \\
\hline
SUN    &	5777 $\pm$ 6  & 0.64  &	4.44  $\pm$ 0.02    &	 0.0                &  0.0 &  4.6 & 25.4 $\pm$ 1.0 &  0.162 $\pm$ 0.005 \\
166	   &  5577 $\pm$ 31  & 0.75 &	4.57  $\pm$ 0.02 &  0.12    $\pm$ 0.02  & 0.18 & $0.00^{+0.84}_{-0.00}$ & 6.23 $\pm$ 0.01 &  0.477 $\pm$ 0.016   \\
1581   &	5927 $\pm$ 61  & 0.57 &  4.45  $\pm$ 0.14 & -0.22	$\pm$ 0.09  & 0.16 & $4.84^{+1.72}_{-2.20}$ & 31.1 $\pm$ 0.1 &  0.150$^{*}$ \\
1835   &	5792 $\pm$ 51  & 0.882 & 4.46 $\pm$	0.09 &	0.17 $\pm$ 0.08	   &  0.22 & $0.00^{+1.76}_{-0.00}$ & 7.8 $\pm$ 0.6 &	0.339 $\pm$ 0.023	 \\
2151   &	5799 $\pm$ 110 & 0.62  & 4.02	$\pm$ 0.17	& -0.14	$\pm$ 0.09	& \ldots	& $6.32^{+0.28}_{-0.24}$ & 23.0 $\pm$ 2.8 & 0.153 $\pm$ 0.015$^{*}$ 	  \\
3651   &	5221 $\pm$	25 & 0.83 & 4.51 $\pm$ 0.02 &	0.16 $\pm$	0.02	&	-0.01 & $11.80$ & 37.0 $\pm$ 1.2 & 0.169 $\pm$ 0.009	 \\
4628   &   4994	$\pm$ 25 &	0.90 &	4.59 $\pm$	0.03 &	-0.19 $\pm$	0.02   &  0.15 & $6.84$ & 38.5 $\pm$ 2.1 &  0.230 $\pm$ 0.026	 \\
10700  & 5341 $\pm$	93 &	0.72 &	4.89 $\pm$	0.21 &	-0.50 $\pm$	0.09	& \ldots & $12.12$ & 34 & 0.171 $\pm$ 0.004  \\
17925  & 5178 $\pm$	100 &	0.86 & 4.53 $\pm$	0.11 &	0.07 $\pm$	0.07	&	0.08 & $0.00^{+1.20}_{-0.00}$ & 7.15 $\pm$ 0.03 & 0.643 $\pm$ 0.045	 \\
18256  & 6535 $\pm$	176	& 0.43 &	4.39 $\pm$ 0.16 &	-0.04 $\pm$	0.13 &  0.24	& 3.2 & 3.65 $\pm$ 0.03 & 0.183 $\pm$ 0.007	 \\
19994  & 6143 $\pm$	92 & 0.531	& 4.16 $\pm$ 0.15  &	0.20 $\pm$	0.07 & 	0.01 & $2.56^{+0.40}_{-0.36}$ & 12.2 &  0.160 $\pm$ 0.003	 \\
20630  & 5708 $\pm$	70 & 0.67	& 4.50 $\pm$ 0.07 &	0.06 $\pm$	0.09 &   0.25 & $0.00^{+2.76}_{-0.00}$ & 9.2 $\pm$ 0.3 &  0.347 $\pm$ 0.021	 \\
20794  &	5467 $\pm$	141 &	0.71 &	4.46 $\pm$ 0.15 &	-0.37 $\pm$	0.11 & \ldots & $12.08$ & 33.19 $\pm$ 3.61 &	0.166$^{*}$\\
22049  &	5101 $\pm$	72	& 0.88 &	4.54 $\pm$	0.13 &	-0.11 $\pm$	0.14 & \ldots & $0.00^{+0.60}_{-0.00}$ & 11.1 $\pm$ 0.1 &	0.505 $\pm$ 0.045	 \\
27383  &	6171	$\pm$ 97 &	0.548 &	4.30 $\pm$	0.12 &	0.07 $\pm$ 0.093 & \ldots & \ldots & \ldots &  \ldots	\\
27524  &	6618	$\pm$ 116 &	0.436 &	4.20 $\pm$	0.04 & 0.13 $\pm$	0.522	& \ldots & \ldots & \ldots  & \ldots \\
30495  &	5833 $\pm$	59	& 0.64	& 4.49 $\pm$	0.09 & 0.005 $\pm$	0.050 &	0.16 & $6.08^{2.12}_{2.20}$ & 11.4 $\pm$ 0.2 &  0.294 $\pm$ 0.015	  \\
33262  &	6158 $\pm$	45 &	0.507 &	4.42 $\pm$	0.02 &	-0.19 $\pm$	0.05 & 0.21 & \ldots & \ldots & 0.272$^{*}$ 	 \\
34411  &	5836 $\pm$	120 &	0.62	& 4.23 $\pm$	0.09 &	0.08	$\pm$ 0.08	& 0.04 & $6.48^{+1.32}_{-1.92}$ & \ldots & 0.145 $\pm$ 0.003 \\
35296  &	6125 $\pm$	48	& 0.53$^{*}$ &	4.28	$\pm$ 0.10 &	0.00 $\pm$	0.06 & 	0.24 & \ldots  & 3.50 $\pm$ 0.01 &  0.317 $\pm$ 0.016 \\
37394  &	5237 $\pm$	58 &	0.84 &	4.52 $\pm$ 0.08 &	0.09 $\pm$	0.08 &  0.06 & $0.00^{+1.36}_{-0.00}$ & 11.49 $\pm$ 0.22 & 0.450 $\pm$ 0.035\\
39587  &   5937 $\pm$	74 & 0.60	& 4.46 $\pm$	0.12 &	-0.04 $\pm$	0.07 &  0.19 & $4.32^{+1.88}_{-2.04}$ & 5.14 $\pm$ 0.01 & 0.319 $\pm$ 0.012\\
44594  &	5817 $\pm$	55  &  0.66 &	4.37 $\pm$	0.04	& 0.14	$\pm$	0.04 &	\ldots & $5.52^{+1.40}_{-1.60}$ & \ldots & 0.155$^{*}$ \\
61421  &	6572  $\pm$	82	& 0.42 &	4.01  $\pm$	0.05 &	-0.03  $\pm$	0.14 &	\ldots & $1.85^{+0.12}_{-0.12}$ & 3 & 0.169 $\pm$ 0.014\\
72905  &  5881  $\pm$	72	& 0.62 &	4.52  $\pm$	0.11 &	-0.07  $\pm$	0.09 & 0.34	& $2.10^{+1.90}_{-1.90}$ & 5.23 $\pm$ 0.02 &  0.360 $\pm$ 0.015\\
81809  &	5630  $\pm$	115 &	0.64 &	3.90  $\pm$ 0.125 &	-0.31	 $\pm$ 0.03 & 0.21 & 3.2 $\pm$ 0.8 & 40.2 $\pm$ 3.0 &  0.172 $\pm$ 0.010\\
82443  & 5311  $\pm$	30 &	0.78$^{*}$ &	4.47	 $\pm$ 0.05 &	-0.09	 $\pm$ 0.10	& 0.10 & \ldots & 5.4 $\pm$ 0.1 &	0.638  $\pm$ 0.050\\
102870	& 6115  $\pm$	57 &	0.55 &	4.15	 $\pm$ 0.11 &	0.15	 $\pm$ 0.07 & 0.10 & $2.96^{+0.24}_{-0.32}$ & \ldots &	0.160 $\pm$ 0.003\\
109358	& 5876  $\pm$	97 &	0.61 &	4.42	 $\pm$ 0.10 &	-0.17  $\pm$	0.11 & 0.15 & $3.68^{+1.64}_{-1.76}$ & \ldots &  0.161 $\pm$ 0.003\\
114710  &	5978  $\pm$	188 &	0.58 &	4.42  $\pm$	0.90	& 0.07  $\pm$	0.09	& 0.25 & $0.00^{+1.12}_{-0.00}$ & 12.3 $\pm$ 1.1 & 0.200 $\pm$ 0.010\\
115383	& 6040	 $\pm$ 80 &	0.58$^{*}$	& 4.24	 $\pm$ 0.15 &	0.14  $\pm$	0.08 & 	0.20 & $1.30^{+1.30}_{-1.30}$ & 3.55 $\pm$ 0.01 & 0.313 $\pm$ 0.016\\
115404	& 4999  $\pm$	50 & 1.03 & 	4.50  $\pm$	0.13 &	-0.16  $\pm$	0.05 & 0.11	& 1.4  & 18.1 $\pm$ 1.3 & 0.523 $\pm$ 0.044\\
115617  &	5550  $\pm$	56	& 0.70 &	4.4  $\pm$	0.1 &	-0.03  $\pm$	0.13 &	0.09 & $8.96^{+2.76}_{-3,08}$ & 26.5 $\pm$ 0.6 & 0.162 $\pm$ 0.005\\
128620A	& 5751  $\pm$	86 &	0.71 &	4.30  $\pm$	0.14 &	0.20  $\pm$	0.06 & \ldots	& $7.84^{+1.08}_{-1.28}$ & 22.5 $\pm$ 5.9 &	0.167 $\pm$	0.032 $^{*}$ \\
129333	& 5751  $\pm$	98  &	0.639 &	4.47  $\pm$	0.10 &	0.01  $\pm$	0.14	& \ldots	& $0.00^{+1.44}_{-0.00}$ & 2.62 $\pm$ 0.01 & 0.543 $\pm$ 0.037\\
133640	& 5695  $\pm$	161	& 0.65 &	4.25  $\pm$	0.09 &	-0.29  $\pm$	0.09 & \ldots	& \dots & \ldots & 0.255 $\pm$ 0.016\\
142373  &	5854  $\pm$	59	& 0.57 &	4.10  $\pm$	0.075	& -0.43  $\pm$	0.09 & 0.33	& $7.76^{+0.36}_{-0.36}$ & 15 & 0.146 $\pm$ 0.003\\
142860	& 6289  $\pm$	103 & 0.50	&  4.23  $\pm$	0.115 &	-0.19  $\pm$	0.07 &  0.2 & $3.56^{+1.20}_{-0.44}$ & \ldots & 0.156 $\pm$ 0.003\\
143761	& 5808	 $\pm$ 61	& 0.60$^{*}$ &	4.12  $\pm$	0.12 &	-0.22	 $\pm$ 0.03	 &	0.16 & $11.04^{+0.88}_{-0.72}$ & 17 &  0.149 $\pm$ 0.004\\
146361	& 5893	 $\pm$ 45	& 0.59 &	4.43  $\pm$	0.13 &	-0.33	 $\pm$ 0.03 &	\ldots & \ldots & \ldots & 0.566 $\pm$ 0.022\\
147513	& 5873	 $\pm$ 64	& 0.644 &	4.52  $\pm$	0.06 & 	0.05	 $\pm$ 0.06 & \ldots & $0.00^{+0.68}_{-0.00}$ & 10.0 $\pm$ 2.0 &  0.23 $\pm$ 0.01 $^{*}$ \\
149661	& 5260	 $\pm$ 47	& 0.78	& 4.17	 $\pm$ 0.09 &	0.03  $\pm$  0.04	& 0.05 	& $0.00^{+4.16}_{-0.00}$ & 21.1 $\pm$ 1.4 & 0.329 $\pm$ 0.027\\
154417	& 6022	 $\pm$ 127	& 0.58$^{*}$	& 4.38	 $\pm$ 0.125 &	-0.005 	 $\pm$ 0.10 & 0.20 & $4.20^{+1.24}_{-1.40}$ & 7.81 $\pm$ 0.06 &	0.268 $\pm$ 0.014\\
173667	& 6363	 $\pm$ 72	& 0.46	& 4.03	 $\pm$ 0.18 &	-0.05  $\pm$ 	0.12 & 0.25	& $3.28^{+0.16}_{-2.12}$ & \ldots &  0.190 $\pm$ 0.001\\
186408	& 5791	 $\pm$ 48	& 0.64	& 4.28	 $\pm$ 0.05 &	0.08   $\pm$	0.05 & 0.05	& $8.36^{+2.92}_{-1.92}$ & 23.8 $\pm$ 1.7 & 0.150 $\pm$ 0.005\\
186427	& 5709	 $\pm$ 55	& 0.66	&  4.34	 $\pm$ 0.05 &	0.07   $\pm$	0.04 & 0.05 & $11.80^{+2.20}_{-2.00}$ & 23.2 $\pm$ 3.0 & 0.150 $\pm$ 0.004\\
187013	& 6312	 $\pm$ 88	& 0.47$^{*}$	& 4.11	 $\pm$ 0.09 &	-0.09  $\pm$ 0.12 & 0.23 & \ldots & 8 &  0.151 $\pm$ 0.004\\
187691	& 6147	 $\pm$ 63	& 0.56$^{*}$	& 4.26	 $\pm$ 0.14 &	0.12 	$\pm$ 0.03 & 0.11 & $3.20^{+0.68}_{-0.40}$ & 10.38 $\pm$ 0.16 & 0.148 $\pm$ 0.005\\
190406	& 5925	 $\pm$ 66	& 0.61	& 4.41	 $\pm$ 0.09 &	0.05 	 $\pm$ 0.04 &	0.08 & $3.16^{+1.84}_{-2.08}$ & 13.9 $\pm$ 0.5 & 0.194 $\pm$ 0.011\\
206860	& 5944	 $\pm$ 87	& 0.58$^{*}$	& 4.48	 $\pm$ 0.12 &	-0.08  $\pm$ 0.07  & 0.12 & $0.00^{+0.88}_{-0.00}$ & 4.85 $\pm$ 0.05 &	0.328 $\pm$ 0.015\\
222368	& 6183	 $\pm$ 87	& 0.50	& 4.13	 $\pm$ 0.17 &	-0.16 	 $\pm$ 0.12	 &  0.14	& $3.44^{+0.24}_{-0.28}$ & \ldots &  0.156 $\pm$ 0.003\\
224930	& 5366	 $\pm$ 180	& 0.67  &	4.42  $\pm$	0.21 &	-0.77	 $\pm$ 0.15 &  0.52	& \ldots & 30.19 $\pm$ 0.95 & 0.184 $\pm$ 0.010
\label{tab1}
\end{longtable}

\subsection{The Line Index Ratios}

The Line Index D is computed as the core-to-wing ratio:
\begin{equation}
\label{def_D}
D =  \frac{\int_{core} E(\lambda)d\lambda}{\int_{cont} E(\lambda)d\lambda}  
\end{equation}
Here, $E(\lambda)$ represents the spectral flux measured at specific wavelengths $\lambda$. The terms {\it{core}} and {\it{cont}} denote the specific wavelength ranges over which the integrals are computed, corresponding to the central region of the spectral line and the nearby continuum, respectively.
These wavelength ranges are provided in Tab.~\ref{tab2}. We stress that for the magnesium lines, the continuum is defined by the average between the two red and blue continua; in particular, the blue continuum of \ion{Mg}{1} coincides with the red continuum of \ion{Mg}{2}. For the \ion{Si}{1} line we decided to consider only the blue continuum, as several IUE stellar spectra appear to degrade at wavelengths greater than 288.3 nm.

\begin{deluxetable*}{cccc}[h!]
\caption{The integral extremes of Eq. \ref{def_D}}
\label{tab2}
\tablewidth{0pt}
\tablehead{
\colhead{Line} & \colhead{Core} & \colhead{Blue Continuum} & \colhead{Red Continuum} \\
       &   (nm)  &    (nm)      &     (nm)         \\
       }
\startdata
\hline
\ion{Mg}{2}  &	279.00 - 281.00  & 276.50 - 277.00  &	283.50 - 284.00 \\
\ion{Mg}{1}  &	285.00 - 285.40  & 283.50 - 284.00  &	286.50 - 287.00 \\
\ion{Si}{1}  &	288.08 - 288.23  & 286.50 - 287.00  &	\ldots  \\
\enddata
\end{deluxetable*}

We computed the values of the indices of \ion{Mg}{2}, \ion{Mg}{1} and \ion{Si}{1} lines for each star in Tab.~\ref{tab1} and for all measurements available. Then, we averaged the different values in time in order to have a unique mean value for each star.

\begin{figure}[h]
\centering
\includegraphics[width=\textwidth]{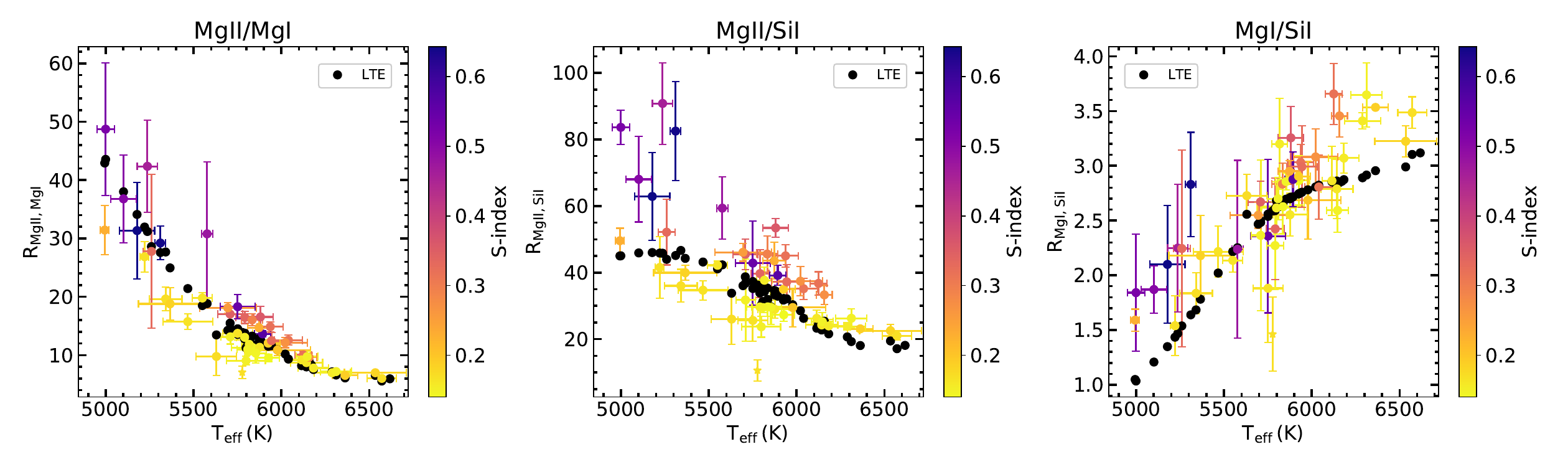}
\caption{The line index ratios R of stars in Tab.~\ref{tab1} plotted as a function of their $T_{eff}$. The three subplots are for: \ion{Mg}{2}/\ion{Mg}{1} (left), \ion{Mg}{2}/\ion{Si}{1} (center) and \ion{Mg}{1}/\ion{Si}{1} (right). The colour map highlights the S-index dependence. The confidence bar for R values represents one standard deviation of the time variability of the single star. 
Black dots represent line index ratios obtained from the LTE synthesis.}
\label{ratio_Teff}
\end{figure}

Figure~\ref{ratio_Teff} shows the line index ratios (R) as a function of the effective temperature for the selected lines. For comparison, the same relation is obtained by calculating LTE synthetic spectra using SPECTRUM, depicted in figure as black filled circles. Briefly, we built a matrix of line index values computed with Kurucz's model with 
$\mathrm{4250 < T_{eff} (K) < 6500}$ with step of 250 K and $\mathrm{1 < log \, g (dex) < 5}$ with step of 0.5;  we then obtained a relation $ \mathcal{R}(T_{eff}, log \, g)$ that we interpolated  by using the stellar parameters in Tab.~\ref{tab1} in order to obtain the R values.
In the same figure, we highlight the dependence on the S-index through a color map. As expected, the ratios also show a dependence on gravity, while there does not appear to be a direct correlation with metallicity (the plots showing the log g and [Fe/H] dependency can be found in Appendix A). The coolest stars are those that show greater variability, linked to the higher S-index value and, thus, to a higher level of magnetic activity.
This is particularly evident in the \ion{Mg}{2}/\ion{Si}{1} ratio, where the trend for these stars significantly deviates from that predicted by the LTE synthesis with models lacking a chromosphere.

Figure~\ref{ratio_Teff} confirms that the ratio between the lines of the same element (i.e. \ion{Mg}{2}/\ion{Mg}{1}) is a better indicator than the other indices. 

\section{Response Functions} 
\label{sec:responses}

We use the Response Function (RF) as a diagnostic to trace changes in temperature, together with theoretical \citep{kurucz} and semi-empirical models \citep{fontenla2011} of stellar atmospheres. The RF gives information about the first-order variations of the emergent intensity due to a perturbation of a given physical parameter \citep[e.g.][]{mein1971,beckers1975, caccin77}. In solar physics, response functions are widely employed to investigate spectral and spectro-polarimetric diagnostics \citep[e.g.][]{cabrerasolana2005, orozco2007, penza2014, quintero2017}, for the characterization of filters \citep[e.g.][]{penza2004, fossum2005, wachter2008, ermolli2010}, and are fundamental tools for spectro-polarimetric inversions \citep[e.g.][]{ruizcobo94,milic2017,li2022,ruizcobo2022}. For the specific case of temperature perturbations, we have:

\begin{equation}
\label{def_RF}
\delta I(\lambda) =  \int_{0}^{\infty} RF(\lambda,\tau) \delta T(\tau) d\tau  
\end{equation}

where T is the temperature, $\lambda$ is the wavelength, and $\tau$ is the optical depth.
We are interested in the response of the ratio of line indices that, after some algebra, can be written as:

\begin{equation}
\label{RF_ij}
RF_{ij} \approx \frac{RF_{core}^{(i)}}{I_{core}^{(i)}} - \frac{RF_{core}^{(j)}}{I_{core}^{(j)}} 
\end{equation} 

where $R_{ij}$ is the ratio between the index of the i-line and the j-line ($R_{ij}= D_{i}/D_{j}$). The steps leading to Eq.~\ref{RF_ij} are detailed in Appendix B.

\begin{figure}[h]
\centering
\includegraphics[width=\textwidth]{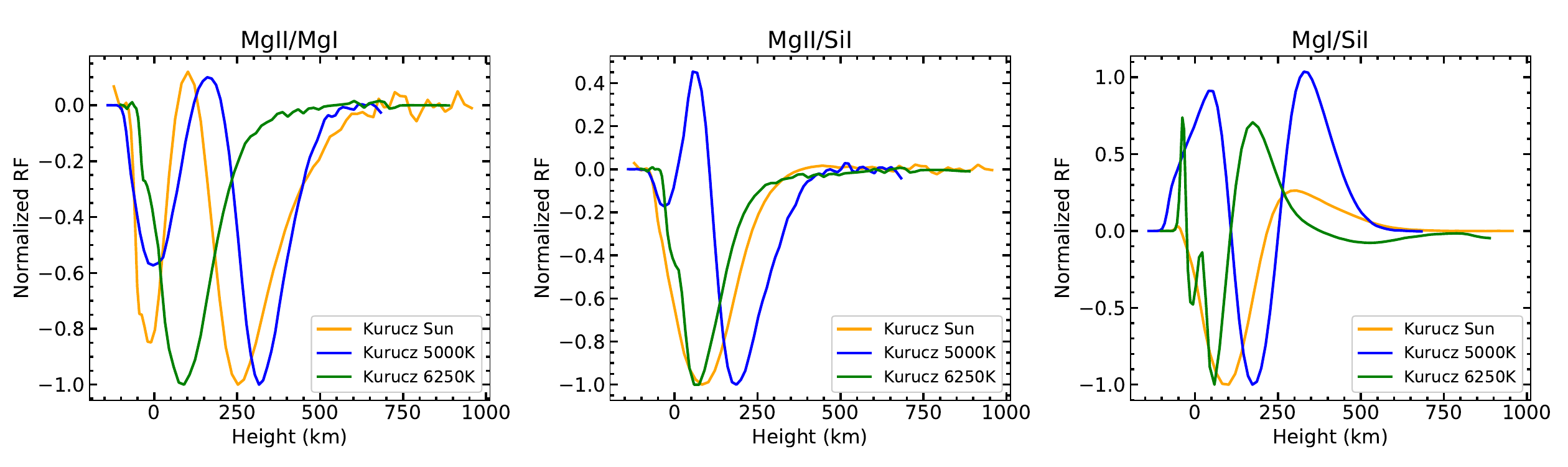}
\caption{Comparison of the temperature response functions (RFs) normalized to their absolute maximum value,  for the three index ratios computed in Non-LTE using the Kurucz solar model (orange line), a Kurucz model with $T_{eff}$ = 5000 K and log g = 4.5 (blue line), and a Kurucz model with $T_{eff}$ = 6250 K and log g = 4.5 (green line). For all three cases, we assumed the solar metallicity (i.e., [Fe/H]=0.00).}
\label{RF_SUN_5000}
\end{figure}

In order to compute the RFs to temperature variations, we employ the numerical approach described in \citet{uitenbroek2006b} by considering different atmospheric models where the temperature profiles are perturbed only in a small interval of height \citep[see also][]{criscuoli2013, criscuoli2023}. 
Specifically, we computed the responses of line index ratios obtained perturbing the temperature of three Kurucz models having solar metallicity (log g = 4.5) and effective temperatures of 5000~K, 5777~K (Sun) and 6250~K. In order to test the goodness of the LTE approximation for computing lines that are typically treated in Non-LTE, the line syntheses, necessary to compute the line indices and their ratios in Eq.~\ref{RF_ij}, were performed using both LTE and Non-LTE. For the Non-LTE syntheses we employed the Rybicki and Hummer (RH) code \citep{uitenbroek2001, criscuoli2019}, since SPECTRUM only allows computations in LTE.   
 The Non-LTE synthesis of the Mg lines was performed using the 76-levels model atom described in \cite{leenaarts2013} and partial frequency redistribution (PRD); the synthesis of the \ion{Si}{1} line was performed using a 15 levels plus continuum atom.
We found that the response functions obtained with the two codes under the different assumptions are very similar. For simplicity, in Fig.~\ref{RF_SUN_5000} we show only the responses of the three indices obtained with RH in Non-LTE. 
These plots provide two pieces of information: 
\begin{itemize} 
\item  for all investigated atmosphere models, the three index ratios are sensitive to temperature variations up to a height (H) of about 600 km, H = 0 km corresponding to the base of the photosphere, where  $\tau_{5000} = 1$
\item for all investigated atmosphere models, the response to the temperature of the index ratio \ion{Mg}{2}/\ion{Mg}{1} is almost always negative, meaning that for variations $\delta T > 0$ the indices decrease; the response of the index ratio  \ion{Mg}{2}/\ion{Si}{1} results negative everywhere in the solar and hotter models, while for colder atmospheres the response present also a positive lobe in the lower photosphere and that explains the change of the slope of synthetic relation in the central plot of Fig.~\ref{ratio_Teff} for $T_{eff} < 5500 K$. Finally, the response of the index ratio \ion{Mg}{1}/\ion{Si}{1} shows a greater dependence on the temperature of the model used. In particular, it 
presents negative and positive lobes with different relative weights for the three models, increasing the positive contribution for cooler stars that overcompensates the negative one. This explains the change of trend at T$_{eff} > 5700$ in the right panel of Fig.~\ref{ratio_Teff}.
\end{itemize}

The different response of the ratios for these models indicates the need to assess the behavior of these lines as the temperature varies, taking into account the different atmospheric models. From this perspective, we note that the ratio between the two Mg lines behaves more consistently than other index ratios.

\subsection{Photospheric or chromospheric indices?}
\label{sec:photorchrom}
\begin{figure}[h]
\centering
\includegraphics[width=\textwidth]{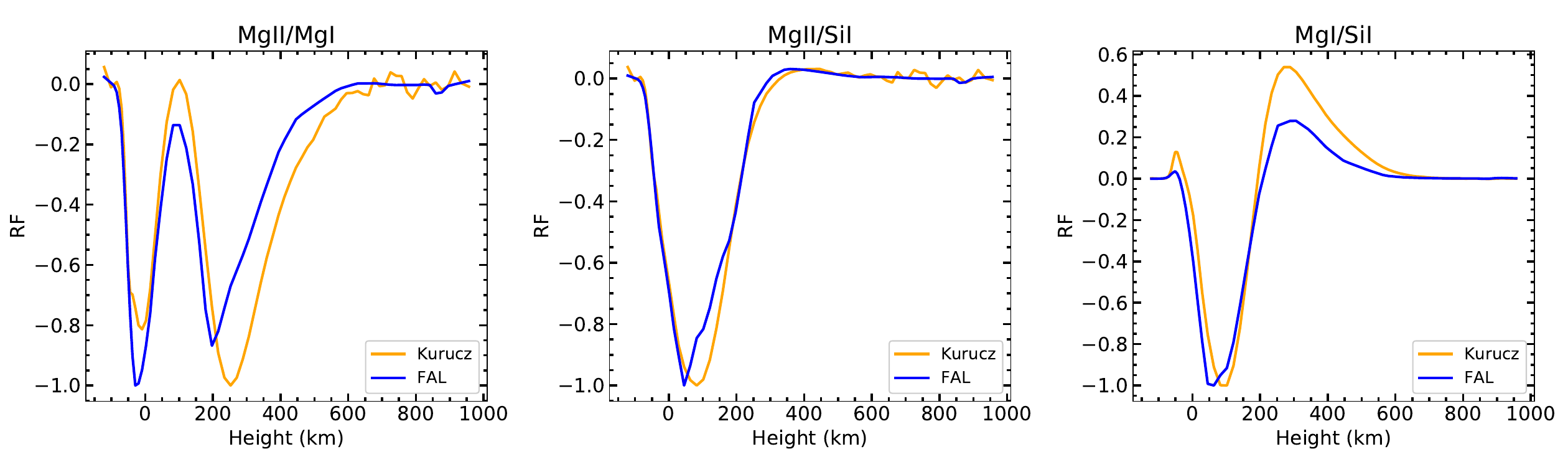}
\caption{Comparison between the temperature RFs of the three index ratios \textbf{computed} by using the FAL model (blue line)  and the Kurucz solar model (orange line).}
\label{RF_LTE_NLTE}
\end{figure}

The RFs shown in Fig.~\ref{RF_SUN_5000} are computed with Non-LTE spectral syntheses that use atmospheric models without temperature chromospheric rise; therefore, it is not surprising that they are consistent with results derived from LTE syntheses performed with SPECTRUM using the grid of the same models. However, we notice that the LTE syntheses obtained with the Kurucz models are also able to reproduce the experimental dependence of the index ratios on $T_{eff}$, particularly the \ion{Mg}{2}/\ion{Mg}{1} ratio. This result is not trivial, as these lines are individually used as indicators of chromospheric activity for single stars. Previous studies showed that collisions significantly influence the \ion{Mg}{2} h and k line formation, making the peak intensities of the core of these lines sensitive to temperatures at their formation heights, which are chromospheric, as demonstrated in several works \citep[e.g.][]{leenaarts2013,leenaarts2013a}. To understand why indices derived from chromospheric lines can serve as proxies for photospheric temperature, we compared RFs computed in LTE using the Kurucz solar model with those obtained in Non-LTE from the quiet Sun model with a chromospheric rise (model 1001) by \citet{fontenla2011} (FAL, hereafter). In both cases, computations are based on RH syntheses. 
The comparison between the RFs obtained with the Kurucz solar model in LTE and the FAL in Non-LTE is shown in Fig.~\ref{RF_LTE_NLTE}.

\begin{figure}[h]
\centering
\includegraphics[scale=0.7]{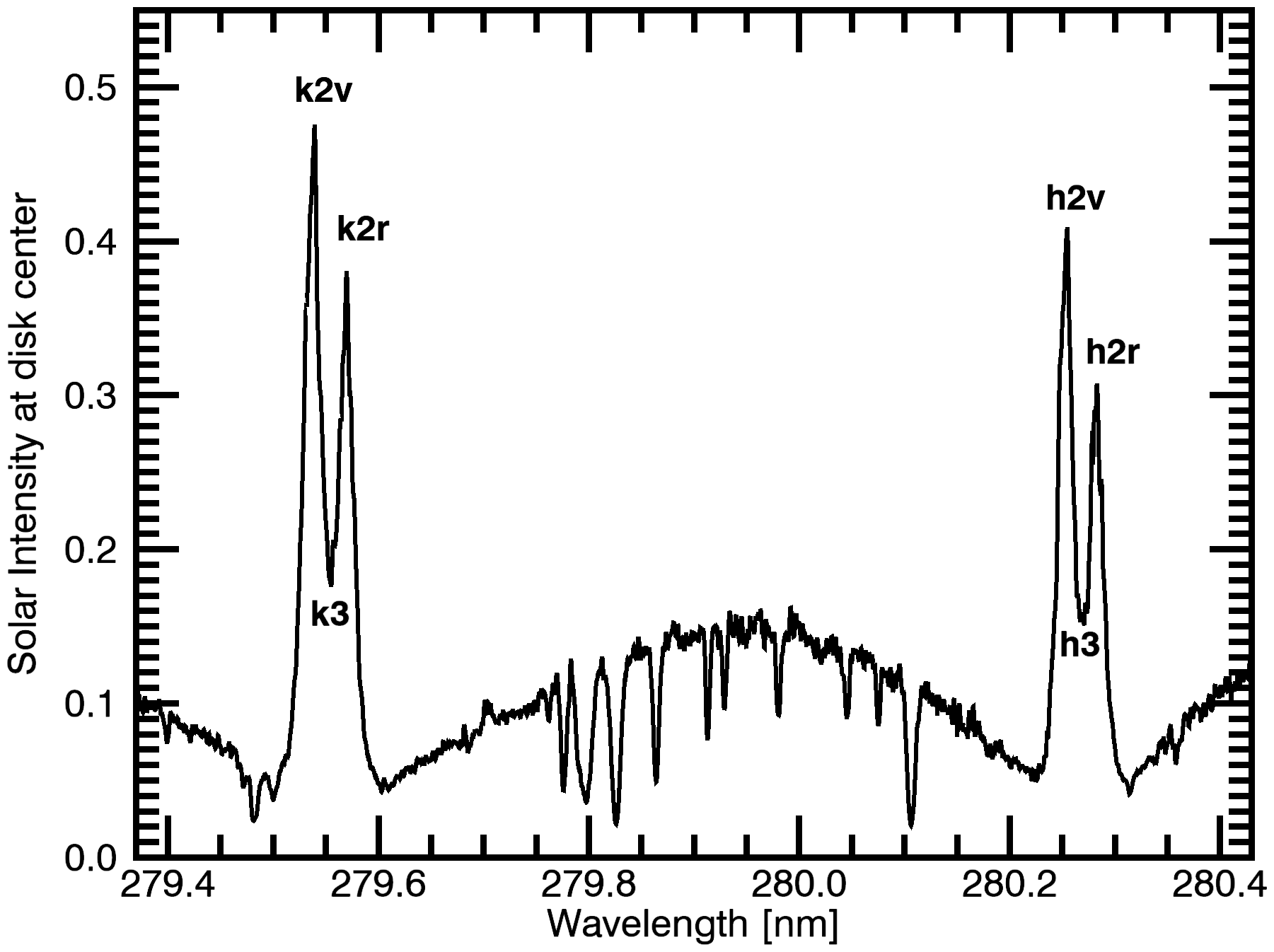}
\caption{Hawaii UV atlas \citep{allen1977} disk center intensity of the solar spectrum around the \ion{Mg}{2} h and k lines, normalized to the nearby continuum. According to conventional nomenclature, the central reversals are denoted with h3 and k3, the violet emission with h2v and k2v, and the red emission with h2r and k2r. }
\label{MgII_core}
\end{figure}

\begin{figure}[h]
\centering
\includegraphics[width=\textwidth]{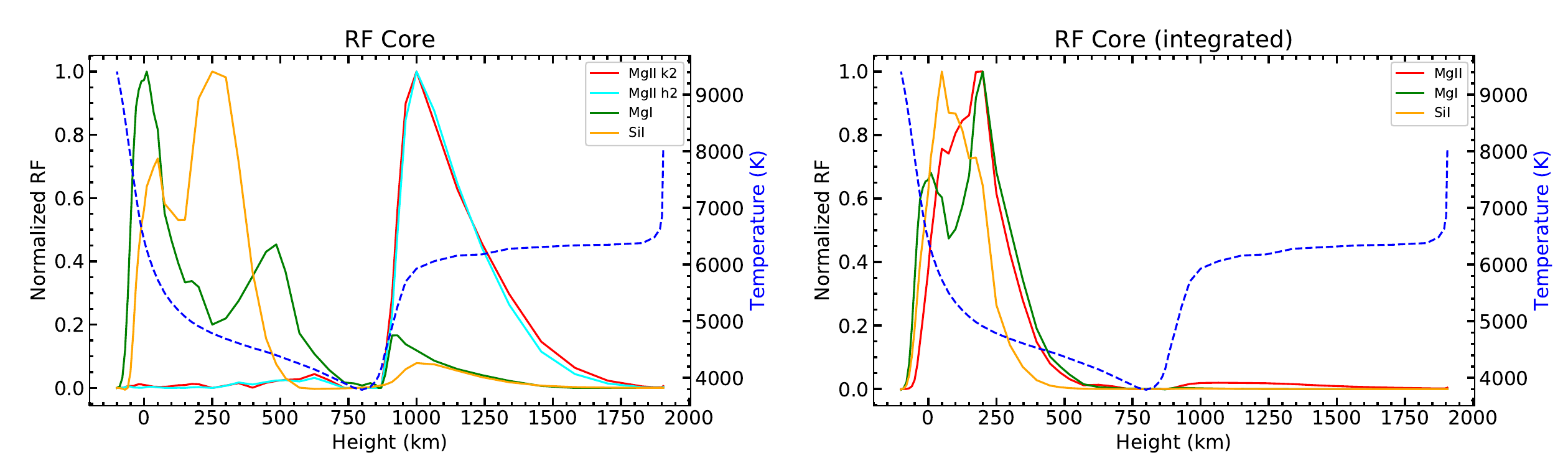}
\caption{Left panel: Temperature Response Functions of the core intensity of the three lines, where for the \ion{Mg}{2} h and k, we show the responses at the h2v and k2v peaks (see Fig.~\ref{MgII_core}). Right panel: Temperature Response Functions of the intensity integrated over the spectral ranges indicated in Sec.~\ref{sec1}. In both panels, the RFs are normalized to their maxima, while the dashed blue line represents the temperature stratification of the FAL model.}
\label{RF_core}
\end{figure}

Although some differences are noticed,  especially for the \ion{Mg}{1}/\ion{Si}{1} ratio, our results clearly indicate that the response functions calculated with the Kurucz models in LTE provide a very good estimate of the temperature sensitivity of these ratios. Comparison of the RFs shown in Fig.~\ref{RF_SUN_5000} and Fig.~\ref{RF_LTE_NLTE} also confirms the good agreement between the LTE and Non-LTE approximations for the Kurucz solar model.
To further understand why the presence of a chromosphere in the model seems to only marginally affect the shape of the RFs,  we investigate the temperature RFs 
as a function of wavelength and height. We focus in particular on the spectral features of the \ion{Mg}{2} h and k lines (illustrated in Fig.~\ref{MgII_core}) and the cores of \ion{Si}{1} and \ion{Mg}{1}, whose corresponding RFs to temperature are shown in Fig.~\ref{RF_core}.
The plots show that especially the individual peaks \ion{Mg}{2} h and k (specifically h2v and k2v), and to a lesser extent, the \ion{Mg}{1} and \ion{Si}{1} cores, respond to temperature variations in the chromospheric layers. In particular, the \ion{Mg}{2} line shows the maximum sensitivity of the h2 and k2 cores at a height of about 1000 km. The \ion{Mg}{1} line core, on the other hand, has a primary maximum in the very lower photosphere, a secondary maximum around 500 km, and a much smaller one around 900 km. Finally, the \ion{Si}{1} line presents two maxima, both in the photosphere, one at 300 km and a second at a height less than 100 km.

However, when we integrate the cores over the wavelength interval used to define the indices (right panel of Fig.~\ref{RF_core}), the resulting responses are predominantly photospheric. This occurs because, on one hand, the responses quickly become photospheric as we move away from the core, with regions such as the area between the h and k cores forming entirely in the photosphere \citep{Uitenbroek1997}. On the other hand, the cores of these lines form in Non-LTE conditions, which makes them less sensitive to local temperature variations \citep{leenaarts2013a}.


\section{$T_{\lowercase{eff}}$ determination by using the $M\lowercase{g}II/M\lowercase{g}I$ ratio}

\begin{figure}[h]
\centering
\includegraphics[scale=0.7]{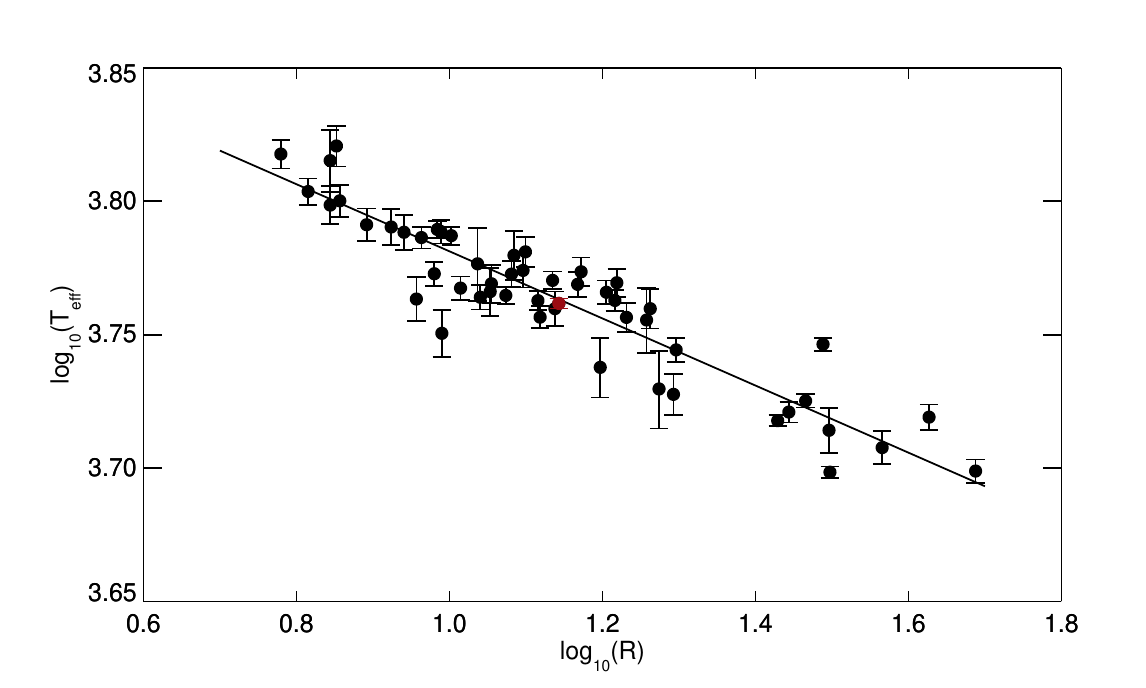}
\caption{Log-log fit between stellar $T_{eff}$ and line index ratio R = \ion{Mg}{2}/\ion{Mg}{1}. The red point represents the Sun.}
\label{Teff_fit}
\end{figure}

We found that the relation between the index ratio R = \ion{Mg}{2}/\ion{Mg}{1} and $T_{eff}$ can be described by a log-log relation:
\begin{equation}
\label{Teff_vs_R}
log(T_{eff}) = a \, log(R) + b 
\end{equation}

where the coefficients of the fit (black line in Figure \ref{Teff_fit}), derived using the data in Fig.~\ref{ratio_Teff}, are:

\begin{eqnarray}
    \nonumber a = -0.126 \pm 0.003 \,\,\,\,\, (log(K))\\
    \nonumber b = 3.907 \pm 0.003 \,\,\,\,\, (log(K)) 
    \label{Teff_cal}
\end{eqnarray}

The corresponding Pearson correlation coefficient of this correlation is r=-0.93. To estimate the statistical significance we performed a t-test and found (t = 17.2) there is a nonzero correlation at a confidence level greater than 99.9\%. We note that the star that deviates the most from this fit is HD 166, that is a young and very active star.

In determining this relationship, we neglected the influence of gravity. In reality, the fit is derived from data that span a range of gravitational values. Furthermore, the \ion{Mg}{2}/\ion{Mg}{1} line ratio shows little variation for stars with effective temperatures above 5200 K, even within the gravity range of $4 < \log g < 5$. 

To evaluate whether R=\ion{Mg}{2}/\ion{Mg}{1} is a good predictor of $T_{eff}$, we computed R for a separate sample of 35 solar-like stars (test-sample) and then applied the log-log relation above to estimate their effective  temperatures. 
Unlike the stars in Tab.~\ref{tab1}, this sample was obtained without requiring the presence of multiple spectra observed at different times. 
The properties of these 35 stars, as reported in the literature, are presented in Tab.~\ref{tab3} alongside our estimates of their effective temperatures ($T_{eff}^{\;\;\;\;(R)}$).

\LTcapwidth=\linewidth
\begin{longtable}{c|c|c|c|c|c|c|c|c|c}
\caption{List of the selected stars and their estimated effective temperature, based on the relation in Eq.\ref{Teff_vs_R}, labeled as $T_{eff}^{\;\;\;\;(R)}$. The other stellar parameters are obtained from the PASTEL and SIMBAD databases, as shown in Tab.~\ref{tab1}.  The B-V values with * are not derived by SIMBAD. Specifically: HD 25998, HD 78366 and HD 97334 by \cite{pizzolato}, HD 26293 by \cite{boro_saikia}. The stellar ages are from \citet{Takeda2007}. $P_{rot}$ values are from \citet{Baliunas1996}, \citet{Hempelmann2016} or \citet{Olspert2018}, except for HD 9826 by \citet{Simpson2010}.
The S-index values with * are not derived by Mount Wilson database, but by \citet{Mittag2018}.}
\label{tab3}
\endfirsthead
\multicolumn{10}{r}{\textit{(To be continued)}}
\endfoot

\multicolumn{10}{l}{\textit{(continued.)}}
\endhead

\endfoot

STAR & $T_{eff}^{\;\;\;\;(R)}$ & $T_{eff}$ &  B-V  & log g    & [$Fe/H$]  &  [$\alpha/Fe$] & Age & $P_{rot}$ & S-index \\
  HD  &  (K) & (K)    &       &  (dex) &   &   & (Gyr) & (days) &       \\
\hline

400	   & 6110 $\pm$  83  & 6173  $\pm$  49 &  \ldots  &	4.12 $\pm$  0.07   &    -0.25  $\pm$  0.07 & 0.30 & $7.52^{+2.92}_{-3.32}$ & \ldots & 0.151 $\pm$  0.004 \\
9826   & 6146 $\pm$  82 & 6154 $\pm$  70 &  0.540  &    4.16 $\pm$  0.13   &     0.068  $\pm$  0.091  & 0.19 & $3.12^{+0.20}_{-0.24}$ & 7.3 $\pm$ 0.04 & 0.154  $\pm$ 0.005 \\ 
10307  & 5978 $\pm$  84 & 5888  $\pm$  74 & 0.620   &   4.33 $\pm$  0.08    &    0.032  $\pm$  0.07 & 0.05 & \ldots & \ldots & 0.151  $\pm$ 0.003 \\
16673  & 6128 $\pm$  83 & 6265  $\pm$  68 &  0.504  &   4.34 $\pm$  0.08    &   -0.002 $\pm$  0.055    & 0.22 & \ldots & 5.98 $\pm$ 0.09 &  0.216 $\pm$ 0.008 \\
22484 & 6055 $\pm$  83  & 5993  $\pm$ 59  &  0.85   &  4.11  $\pm$ 0.15     & -0.08 $\pm$ 0.06    &  0.09 & $5.64^{+0.44}_{-1.72}$ & \ldots & 0.146 $\pm$ 0.003\\
25998 & 6007 $\pm$  84  & 6356  $\pm$ 160   & 0.52$^{*}$   &     4.56  $\pm$  0.28      & 0.15  $\pm$ 0.18  & 0.28 & \ldots & 2 &  0.286 $\pm$ 0.014\\
26923 & 5817 $\pm$  86 & 5985  $\pm$ 69    &  \ldots     &   4.45  $\pm$ 0.04      &  0.002  $\pm$ 0.06  &  0.18 & \ldots & 10.6 $\pm$ 0.3 & 0.283 $\pm$ 0.012\\
27406 & 5911 $\pm$  85 & 6109  $\pm$ 103 &  0.57$^{*}$      &   4.25 $\pm$ 0.11       &  0.12 $\pm$ 0.06   & 0.13 & \ldots & \ldots & 0.289 $\pm$ 0.001$^{*}$  \\
27836 & 5936 $\pm$  84  & 5760  $\pm$ 33  &  0.634  & 4.3  & 0.16    & 0.24 & \ldots & \ldots & 0.345 $\pm$ 0.011 $^{*}$ \\
27859  & 5595 $\pm$  87 & 5891 $\pm$ 102  &  0.592  &  4.40 $\pm$ 0.08  & 0.12 $\pm$ 0.06 &  0.11 & \ldots & \ldots & 0.296 $\pm$ 0.012 $^{*}$\\
28068 & 5659 $\pm$  87 & 5757 $\pm$ 199  &  0.64  &  4.41 $\pm$ 0.08     &  0.07 $\pm$ 0.07  & 0.21 & \ldots & \ldots & 0.329 $\pm$ 0.032$^{*}$  \\
28205  & 5985 $\pm$  84 & 6220 $\pm$ 42    &  0.545 &  4.305 $\pm$  0.005   &  0.142 $\pm$ 0.05 & 0.23 & \ldots & \ldots & 0.238 
$\pm$ 0.002 $^{*}$ \\
28344 & 5682 $\pm$ 87   & 5921  $\pm$ 230 &  0.619 &  4.43  $\pm$ 0.06 & 0.13  $\pm$ 0.09 & 0.16 & \ldots & \ldots & 0.297 $\pm$ 0.020$^{*}$ \\
28992  & 5345  $\pm$  90 & 5882 $\pm$ 68  & 0.632  & 4.42 $\pm$ 0.10  & 0.14 $\pm$ 0.04  &  0.14 & \ldots & \ldots & 0.301 $\pm$ 0.007 $^{*}$ \\
33256 & 6563 $\pm$  78  & 6376    $\pm$ 88 &  0.427 &   3.95  $\pm$ 0.18     & -0.37 $\pm$ 0.11     &  0.16 & \ldots & \ldots & 0.153 $\pm$ 0.002 \\
43042 & 6342 $\pm$  80 & 6508   $\pm$ 59  &  \ldots     &   4.26  $\pm$ 0.05    &  0.05 $\pm$ 0.05    & 0.13 & $2.28^{+0.32}_{-0.36}$ & \ldots &  0.163   $\pm$ 0.003\\
43318 & 6421 $\pm$  79  & 6256    $\pm$ 60 &  0.50  &   3.88  $\pm$ 0.16    &  -0.17 $\pm$ 0.05    &  0.17 & \ldots & \ldots & \ldots \\
48682 & 6062 $\pm$ 83  & 6052   $\pm$ 125 &  \ldots    &    4.35  $\pm$ 0.19    & 0.11  $\pm$ 0.04  & 0.14  & $4.00^{+3.20}_{-0.92}$ & \ldots & 0.151  $\pm$ 0.004 \\
76932 & 6150 $\pm$  82 & 5869  $\pm$  69  &  0.53 &   4.04 $\pm$ 0.23    &  -0.91 $\pm$ 0.09  &   \ldots & \ldots & \ldots & \ldots \\
78366 & 5869 $\pm$ 85  & 5995   $\pm$ 41   &  0.60   &  4.50 $\pm$ 0.1   &  0.04 $\pm$ 0.03   &  0.16 & $0.00^{+0.68}_{-0.00}$ & 9.52 $\pm$ 0.08 & 0.245  $\pm$ 0.019 \\
82885 & 5453 $\pm$ 89  & 5508  $\pm$  95  &  0.77 &   4.44  $\pm$ 0.13    &  0.31 $\pm$ 0.09  &  -0.03  & \ldots & 17.88 $\pm$ 0.18 & 0.288  $\pm$ 0.025 \\
84737  & 5897 $\pm$  85  & 5896  $\pm$  46  &  0.62  &  4.13  $\pm$ 0.11   & 0.093 $\pm$ 0.06   & 0.07  & $4.08^{+0.36}_{-0.28}$ & 4.30 $\pm$ 0.02 &  0.136  $\pm$ 0.004\\
89449  & 6298 $\pm$  80 & 6472   $\pm$ 55 &    \ldots  &  4.13  $\pm$ 0.01   & 0.11  $\pm$ 0.01  & 0.22 & \ldots & \ldots & 0.414  $\pm$ 0.015 \\
97334  & 5691 $\pm$  87 & 5867   $\pm$ 52   &   0.60 & 4.36 $\pm$ 0.10  & 0.05 $\pm$ 0.03 &  0.18 & $0.00^{+2.92}_{-0.00
}$ & 7.93 $\pm$ 0.05 & 0.333  $\pm$ 0.017\\
101501 & 5403 $\pm$  89  & 5465  $\pm$ 155 &   0.74  & 4.54 $\pm$ 0.05  & -0.06 $\pm$ 0.09 &  0.07 & 11.32 & 15.9 $\pm$ 0.2 & 0.303   $\pm$ 0.024\\
103095  & 5389 $\pm$  89  & 5052   $\pm$  70 &  0.75  &  4.57   $\pm$ 0.24   &  -1.34  $\pm$ 0.12 & 0. 58 & $0.00^{+2.44}_{-0.00}$ & 34.03 $\pm$ 0.68 & 0.184  $\pm$ 0.011 \\
106516 & 6220 $\pm$  82 & 6157 $\pm$  175  &  0.46 & 4.36 $\pm$  0.20  & -0.70 $\pm$ 0.19 & \ldots & \ldots & 6.63 $\pm$ 0.04 &  0.208  $\pm$ 0.008\\
114378 & 6295 $\pm$  81 & 6382 $\pm$ 25 & 0.572 &  4.18  $\pm$ 0.11  &  -0.19  $\pm$ 0.06   & \ldots & \ldots & 4.39 $\pm$ 0.02 & 0.241  $\pm$ 0.007\\
115043 & 5629 $\pm$  87 & 5749    $\pm$ 260  & 0.61  & 4.47     &  -0.06  $\pm$ 0.05  &   0.23 & \ldots & 5.5 $\pm$ 0.1 & 0.321  $\pm$ 0.018\\   
120136 & 6200 $\pm$  82 & 6479 $\pm$ 110 &  0.49 & 4.32 $\pm$ 0.25  & 0.28 $\pm$  0.10    &  0.30 & $1.64^{+0.44}_{-0.52}$ & 3.07 $\pm$ 0.06 & 0.188  $\pm$ 0.006 \\
141004 & 5906 $\pm$ 85  & 5908   $\pm$  76  & 0.61 & 4.17 $\pm$  0.10  & -0.007 $\pm$  0.050 &  0.14 & $6.32^{+0.88}_{-1.56}$ & 26 & 0.156  $\pm$ 0.006\\
152391 & 5370 $\pm$  89 & 5452  $\pm$ 46 & 0.76   & 4.49  $\pm$ 0.09  & -0.03  $\pm$ 0.07 &  0.12 & \ldots & 10.62 $\pm$ 0.13 & 0.391  $\pm$ 0.036 \\ 
185144 & 5554 $\pm$  88 & 5289   $\pm$ 90  &  0.78  &  4.49   $\pm$ 0.13    &  -0.22  $\pm$ 0.10  & 0.03  & \ldots & 27.7 $\pm$ 0.8 & 0.218  $\pm$ 0.020 \\
217014 & 5787 $\pm$  86 & 5766   $\pm$   59 & 0.7  &  4.31 $\pm$  0.14 & 0.18 $\pm$ 0.06  & 0.03  & $6.76^{+1.64}_{-1.48}$ & 38.0 $\pm$ 0.6 & 0.149 $\pm$ 0.004 \\
284253 & 5393 $\pm$  89 & 5331   $\pm$ 93   & 0.81  &   4.505   $\pm$ 0.009 &  0.12  $\pm$ 0.02 & 0.04 & \ldots & \ldots &  0.412 $\pm$ 0.016 $^{*}$ \\
%
\end{longtable}

\begin{figure}[h]
\centering
\includegraphics[width=0.5\textwidth]{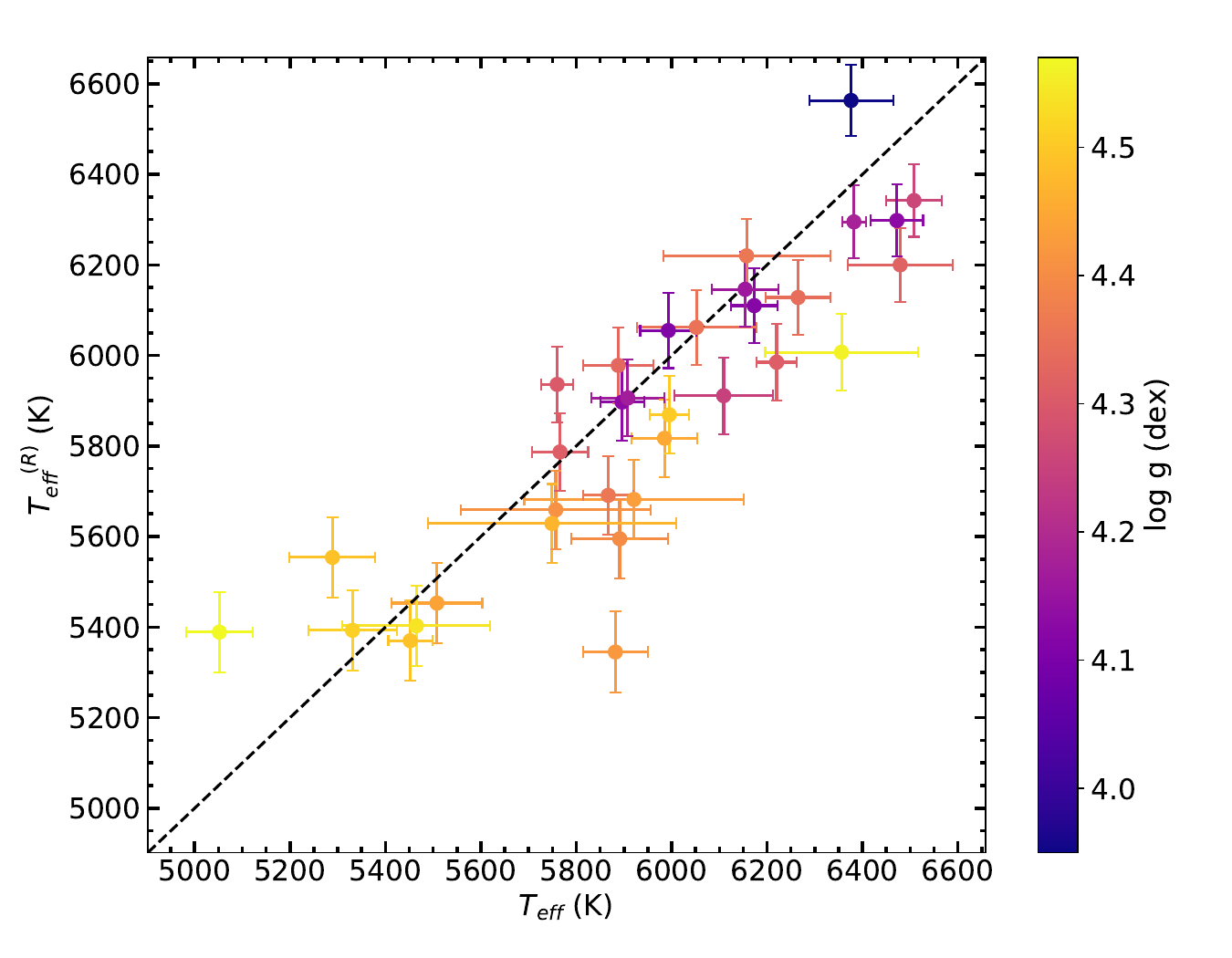}
\includegraphics[width=0.5\textwidth]{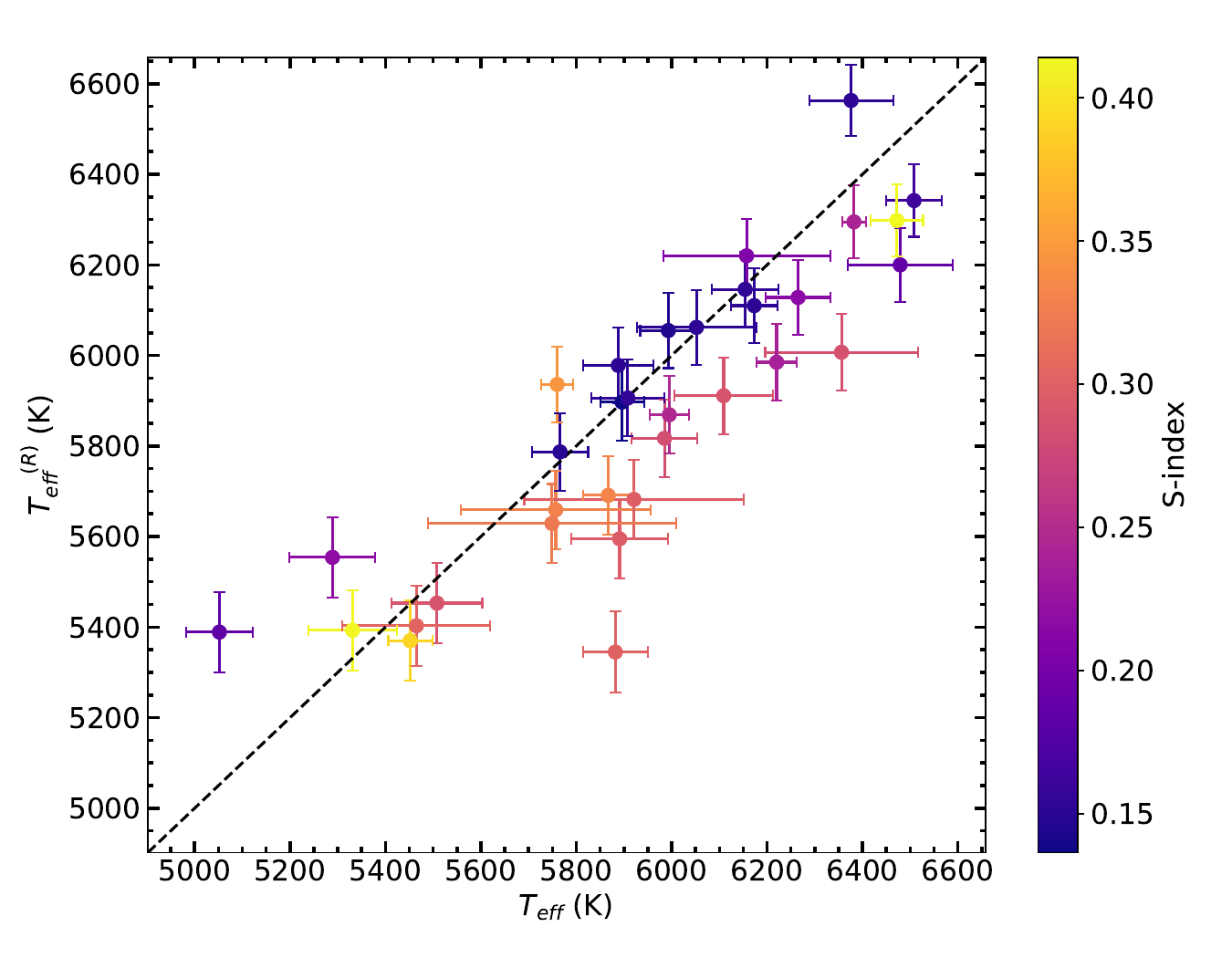}
\includegraphics[width=0.5\textwidth]{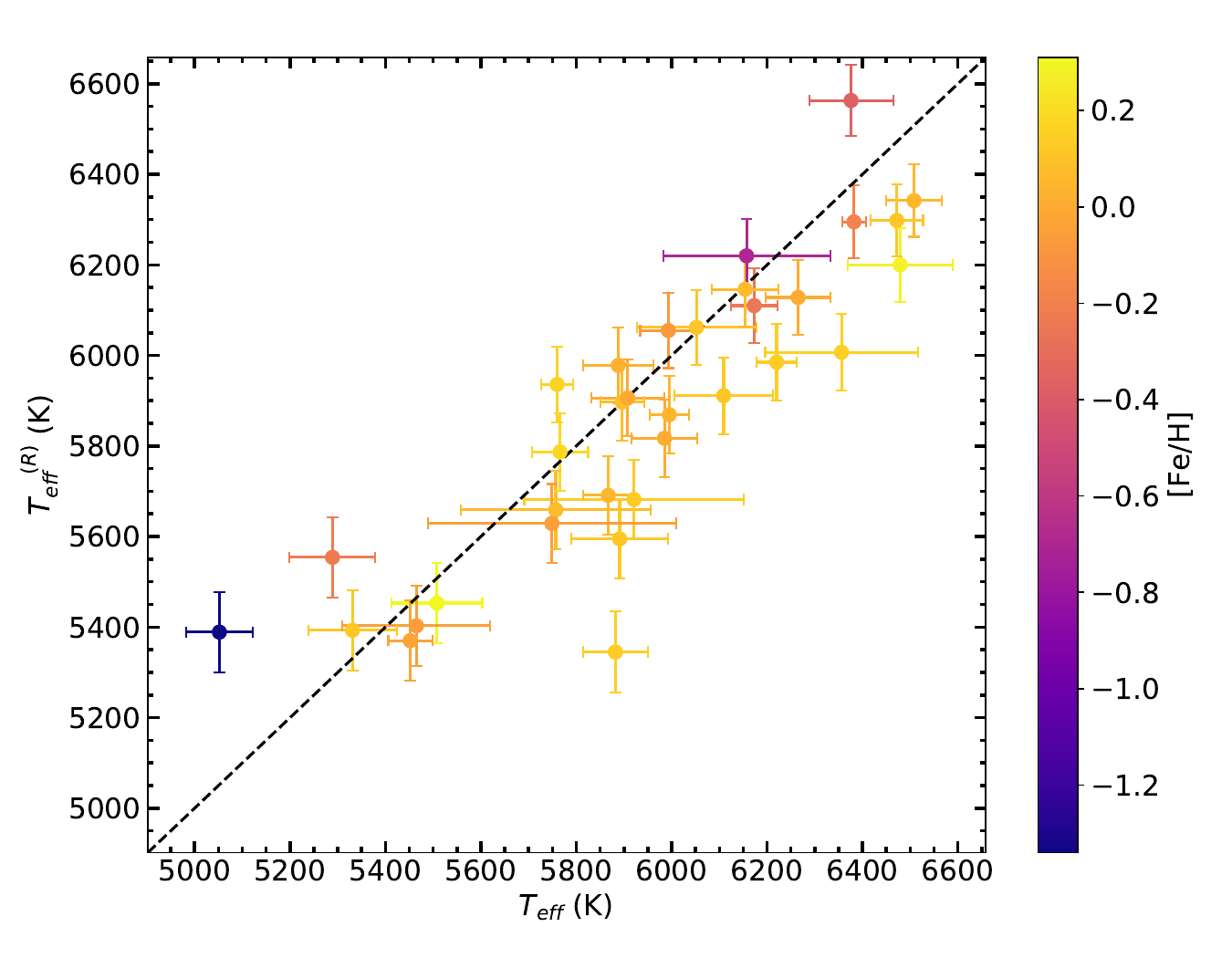}
\caption{The effective temperatures obtained in this work ($T_{eff}^{\;\;\;\;(R)}$, y-axis), for the stars in Tab.~\ref{tab3}, against the literature values (x-axis). In all panels the dashed black line indicates the bisector, while the color map shows the dependence on log g (top panel), S-index (middle panel) and [Fe/H] (bottom panel).}
\label{Teff_test}
\end{figure}

In Fig.~\ref{Teff_test} we plot our effective temperature determinations ($T_{eff}^{\;\;\;\;(R)}$) versus the average values from the literature ($T_{eff}$), highlighting their dependence on $\log g$, S-index and metallicity using a color coding. Except for a few stars in the sample, the effective temperature determinations appear to agree with the values reported in the literature. To quantify this agreement, we performed a Wilcoxon signed-rank test. The Wilcoxon signed-rank test is a non-parametric statistical method used for hypothesis testing. It is applied either to evaluate the location of a population based on a sample of data or to compare the locations of two populations using two matched samples. For the sample of stars shown in Fig.~\ref{Teff_test}, the test statistic ($W$) is calculated as $W$=210.
This value is compared to the critical value from the Wilcoxon Signed-Rank Test Critical Values
Table (two-tailed) corresponding to $n$ = 35 (the number of stars in the sample) and
$\alpha$ =0.05 (95\% confidence level). The critical value for these parameters is 195.
Since the calculated $W$ (210) is greater than the critical value (195), we conclude that there
is no significant difference between the two population medians at the 95\% confidence level. That is, our results are in good agreement with previous studies.

The reliability of the empirically derived log-log relation to estimate the effective temperature will be examined further in the next section.


\section{Discussion and conclusions}\label{sec:sec6}

In this study, we examined the dependence of the ratios of core-to-wing indices on stellar parameters, focusing on three mid-UV lines: \ion{Mg}{2} 280.0 nm, \ion{Mg}{1} 285.2 nm, and \ion{Si}{1} 288.16 nm. We analyzed spectra from 52 solar-like stars in the IUE database and explored how the derived line index ratios relate to stellar fundamental parameters. Our findings reveal a strong dependence on effective temperature and a secondary dependence on gravity, suggesting that the UV line index ratios investigated here are effective indicators of photospheric temperatures. Among the indices examined, the \ion{Mg}{2}/\ion{Mg}{1} ratio exhibits the least scatter and the clearest trend with effective temperature. Furthermore, this ratio shows the best agreement with LTE synthesis based on Kurucz's atmospheric models for solar-like stars, making it the most reliable indicator of photospheric conditions.

To gain insight into our findings, we investigated the RFs to temperature perturbations applied to models with and without a chromosphere. We confirm the chromospheric sensitivity of the individual peaks \ion{Mg}{2} h and k, while the cores of \ion{Mg}{1} and \ion{Si}{1} are more sensitive to temperature perturbations at lower heights in the atmosphere.
However, we showed that the responses of the cores of the three lines shift toward the photosphere when integrating over the wavelength ranges used to define the indices described in Tab.~\ref{tab2}. Consequently, also the line indices derived from the three lines and their ratios show the highest sensitivity to temperature perturbations in the photosphere. This result aligns with results previously found for the \ion{Ca}{2} H and K lines, whose cores form in the chromosphere, but whose responses become photospheric when observed with broadband filters \citep{ermolli2010,murabito2023} typically employed for monitoring solar activity.
However, the line ratios we considered exhibit different responses to temperature variations. Specifically, \ion{Mg}{2}/\ion{Mg}{1} and \ion{Mg}{2}/\ion{Si}{1} decrease with positive temperature perturbations, whereas \ion{Mg}{1}/\ion{Si}{1} shows the opposite trend. In Sec.~\ref{sec:photorchrom} we showed that the exact sign, shape, and amplitude of the RFs of the line indices and their ratios depend on the atmosphere model; however, responses obtained in Non-LTE using an atmosphere with a chromosphere are well matched by responses obtained using LTE approximation with models without a chromosphere.

Supported by the results obtained from the analysis of synthetic spectra, which indicate that all the analyzed index ratios are sensitive to photospheric perturbations, we derived an empirical log-log relation between the observed index ratio \ion{Mg}{2}/\ion{Mg}{1} and T$_{eff}$, and used this relation to estimate the effective temperature of a second sample of 35 solar-like stars (test-sample) from the IUE database. 
We found that the estimated temperatures agree with the ones provided in the literature at a 95\% confidence level.  

However, examining the plots in Fig.~\ref{Teff_test}, it is clear that the effective temperatures derived from Eq.~\ref{Teff_vs_R} tend to be slightly underestimated for stars with high magnetic activity. 
Indeed, all stars listed in Tab.~\ref{tab3} with an S-index $\gtrsim$ 0.22 exhibit $T_{eff}$ estimates lower than those in the literature. This discrepancy may stem from the use of a single spectrum, which could have led to higher R values than the average, thereby underestimating $T_{eff}$. On the other hand, low metallicity appears to significantly contribute to overestimating $T_{eff}$. A notable example is HD 103095, an extremely metal-poor star ([Fe/H] $\simeq$ -1.34), which shows the largest $T_{eff}$ overestimation. 
Additionally, the results in the top panel of Fig.~\ref{Teff_test} suggest that for stars with surface gravity values log g $\lesssim$ 4.0 or log g $\gtrsim$ 4.5 the proposed method will most likely provide $T_{eff}$ estimates that deviate from values reported in the literature. 
The use of line depth ratios as $T_{eff}$ calibrators is already well established, but so far it has been mostly limited to lines in the visible and infrared ranges. 
Here, we demonstrate that it is possible to extend this approach to the mid-UV range, with the advantage of using the same lines as diagnostics of stellar chromospheres and stellar fundamental parameters. 
In particular, Fig. \ref{RF_core} illustrates how different atmospheric heights are sampled depending on the width of the spectral range of integration around the line cores. The capability to investigate the height dependence of temperature, from the photosphere to the upper chromosphere, using the same spectral line, exists by progressively refining the core selection. This approach will be the focus of a future study.
Another possible extension of this work could involve new calibrations using stellar data from other UV spectral observations, such as those from SOLSTICE \citep{snow}, Hubble Space Telescope \citep[e.g.][]{sahu}, FUSE and CUTE, as well as from future instruments such as MAUVE, HWO and MANTIS. However, in the context of UV spectroscopy, the game changer appears to be UVEX\footnote{https://www.uvex.caltech.edu}\citep{kulkarni2021}, the new NASA Medium Explorer mission to explore the ultraviolet sky, which is planned to be launched in 2030. Indeed, this satellite will cover the entire sky simultaneously in the FUV and in the NUV bands. This means a unique opportunity to use UV spectral diagnostics to investigate a wide range of young and old stellar tracers in the Milky Way and in nearby stellar systems (globulars, dwarf galaxies).

In light of this, our next objectives are twofold: first, to analyze the behavior of the studied lines in the context of solar variability, and second, to extend the approach by incorporating different pairs of UV lines for M-type stars. This is particularly important because M dwarfs are key targets in the search for exoplanets due to their abundance, favorable conditions for detecting transiting planets, and the significant impact of their activity on planetary habitability. The applicability of the method presented in this work to stars significantly different from the Sun, in terms of gravity, abundance, or high magnetic activity, will require a more detailed investigation. This should be done by exploring different atmospheric models and analyzing additional data to assess the effect of such conditions on these spectral lines. 



\begin{acknowledgments}

The authors thank the anonymous referee for the helpful suggestions and advice provided.
The National Solar Observatory is operated by the Association of Universities
for Research in Astronomy (AURA), Inc. under a cooperative
agreement with the National Science Foundation.  It is a pleasure to thank F. Verrecchia and J. Mullen for useful discussions concerning current and future UV photometric and spectroscopic observing facilities. This investigation was partially supported by Project PRIN MUR 2022 (code 2022ARWP9C) “Early Formation and Evolution of Bulge and HalO (EFEBHO)” (PI: M. Marconi), funded by European Union—Next Generation EU, by the project “Realizzazione di un’Infrastruttura HW e SW (IHS) presso il CGS/Matera (CUP F83C22002460005), in attuazione del Piano Operativo del sub-investimento M1C2.I4.4 “In-Orbit Economy - SST – FlyEye” del PNRR-FC”, by the MELODY project (grant agreement No. SOE 0000119, CUP E53C22002450006) under the “Funding projects presented by young researchers” initiative, supported by the European Union - NextGenerationEU", and by the Large grant INAF 2023 MOVIE (PI: M. Marconi). We
made much use of NASA’s Astrophysics Data System
bibliographic services.  

The IUE data presented in this article were obtained from the Mikulski Archive for Space Telescopes (MAST) at the Space Telescope Science Institute. The specific observations analyzed can be accessed via \dataset[https://doi.org/10.17909/a2vq-qf63]{https://doi.org/10.17909/a2vq-qf63}.

\end{acknowledgments}

\section{Appendix A}
\label{appendix_a}

\begin{figure}[h!]
    \centering
    \includegraphics[width=\textwidth]{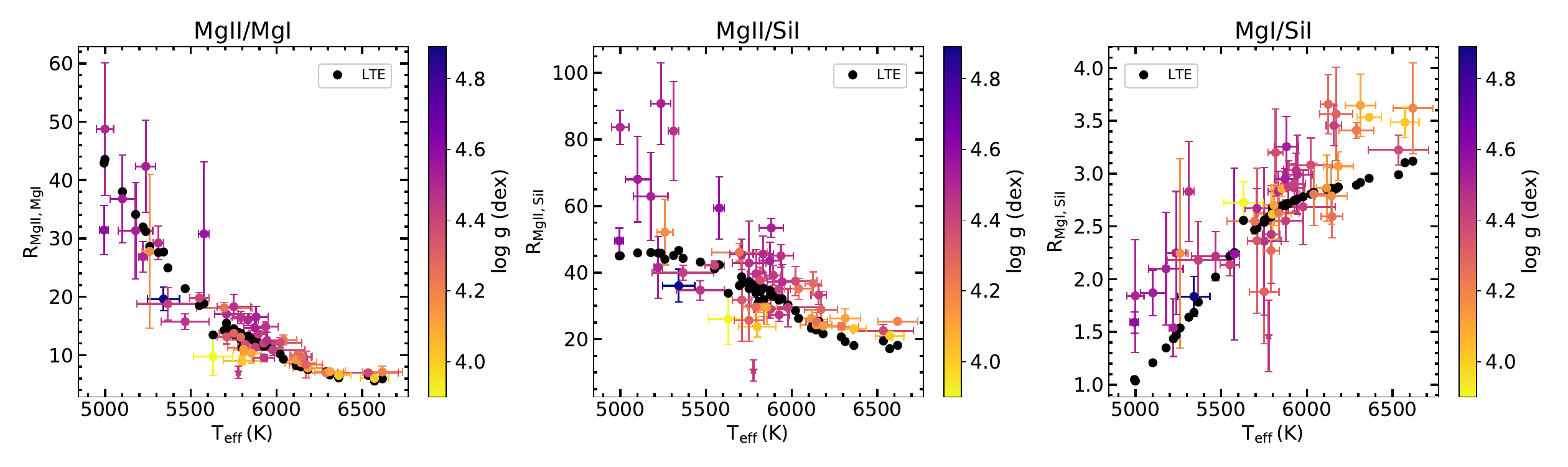}
    \caption{
    Same as Fig~\ref{ratio_Teff}, but with the color map highlighting the log g dependence.}
    \label{fig_2_log_g}
\end{figure}

\begin{figure}[h!]
    \centering
    \includegraphics[width=\textwidth]{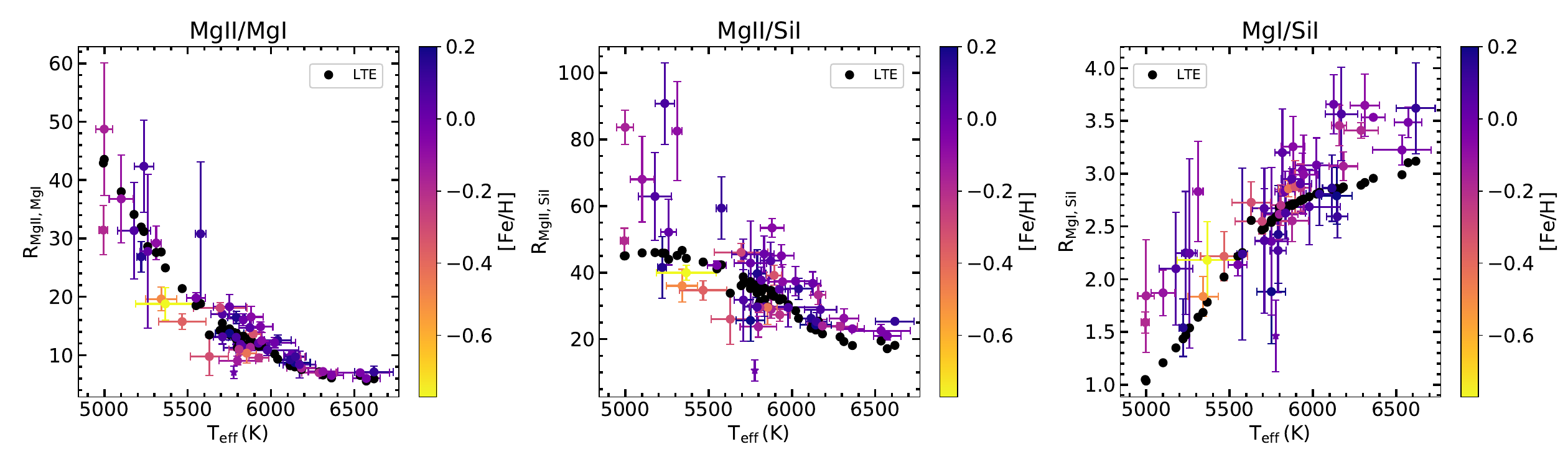}
    \caption{Same as Fig.~\ref{ratio_Teff}, but with the colour map highlighting the [Fe/H] dependence.}
    \label{fig_2_fe_h}
\end{figure}

We report here the color maps showing the dependence of the line index ratios  \ion{Mg}{2}/\ion{Mg}{1}, \ion{Mg}{2}/\ion{Si}{1} and \ion{Mg}{1}/\ion{Si}{1}  on log g (Fig.~\ref{fig_2_log_g}) and [Fe/H] (Fig. \ref{fig_2_fe_h}). These figures are the same as in Fig.~\ref{ratio_Teff}, except for the different color-coded dependence.

\section{Appendix B}
\label{appendix_b}

We detail here the algebraic steps that lead to the final expression of the Response Function of the line index ratios reported in Eq. \ref{RF_ij}. To simplify the notation, we omit the dependencies on $\lambda$ and $\tau$, so we can write the variations for D as:
\begin{equation}
\label{}
\frac{\delta D}{D} =  \frac{\delta I_{core}}{I_{core}} -  \frac{\delta I_{cont}}{I_{cont}} = \frac{\int_{0}^{\infty} RF_{core} \delta T d\tau}{I_{core}} - \frac{\int_{0}^{\infty} RF_{cont} \delta T d\tau}{I_{cont}} = \frac{\int_{0}^{\infty} (RF_{core}- D ^{.} RF_{cont}) \delta T d\tau}{I_{core}}
\end{equation}
Then, the Response Function for the single line index D results:
\begin{equation}
\label{RF_D}
RF_{D} =  \frac{D}{I_{core}} (RF_{core} - D RF_{cont})  
\end{equation}
In analogous way, variations of the ratio ($R_{ij} \equiv D_i/D_j$) of two different indices $i, j$ result:
\begin{equation}
\frac{\delta R_{ij}}{R_{ij}} = \frac{\delta D_i}{D_i} -  \frac{\delta D_j}{D_j} = \frac{\int_{0}^{\infty} (RF_{D_i}- R_{ij} ^{.} RF_{D_j}) \delta T d\tau}  {D_1} 
\end{equation}
With simple algebraic passages, we obtain:
\begin{equation}
\frac{\delta R_{ij}}{R_{ij}} = \int_{0}^{\infty}  \left[ \frac{RF_{core}^{(i)}}{I_{core}^{(i)}} - \frac{RF_{cont}^{(i)}}{I_{cont}^{(i)}} - \frac{RF_{core}^{(j)}}{I_{core}^{(j)}} + \frac{RF_{cont}^{(j)}}{I_{cont}^{(j)}} \right] \delta T d\tau  
\end{equation} 
Because the spectral lines are very close to each other, we can assume that the continuum intensities and their response functions are the same, so that the relation above can be rewritten as:

\begin{equation}
\frac{\delta R_{ij}}{R_{ij}} \approx \int_{0}^{\infty}  \left[ \frac{RF_{core}^{(i)}}{I_{core}^{(i)}} - \frac{RF_{core}^{(j)}}{I_{core}^{(j)}} \right] \delta T d\tau  
\end{equation}

\bibliography{main.bib}{}
\bibliographystyle{aasjournal}



\end{document}